\documentclass[twocolumn,showpacs,showkeys,superscriptaddress,preprintnumbers,amsmath,amssymb,prl]{revtex4}
\usepackage{graphicx}
\usepackage{dcolumn}
\usepackage{bm}
\usepackage{amsmath}
\begin{document}

\title{Nonlinear Force Propagation During Granular Impact}

\author{Abram H. Clark}
\author{Alec J. Petersen}
\affiliation{Department of Physics \& Center for Nonlinear and Complex Systems, Duke University, Durham, North Carolina 27708, USA}
\author{Lou Kondic}
\affiliation{Department of Mathematical Sciences, New Jersey Institute of Technology, Newark, New Jersey 07102, USA}
\author{Robert P. Behringer}
\affiliation{Department of Physics \& Center for Nonlinear and Complex Systems, Duke University, Durham, North Carolina 27708, USA}

\begin{abstract}

We experimentally study nonlinear force propagation into granular material during impact from an intruder, and we explain our observations in terms of the nonlinear grain-scale force relation. Using high-speed video and photoelastic particles, we determine the speed and spatial structure of the force response just after impact. We show that these quantities depend on a dimensionless parameter, $M'=t_c v_0/d$, where $v_0$ is the intruder speed at impact, $d$ is the particle diameter, and $t_c$ is the collision time for a pair of grains impacting at relative speed $v_0$. The experiments access a large range of $M'$ by using particles of three different materials. When $M' \ll 1$, force propagation is chain-like with a speed, $v_f$, satisfying $v_f \propto d/t_c$. For larger $M'$, the force response becomes spatially dense and the force propagation speed departs from $v_f\propto d/t_c$, corresponding to collective stiffening of a strongly compressed packing of grains.

\end{abstract}
\date{\today}

\maketitle

The topic of impact into a granular medium is of considerable recent interest~\cite{Allen1957,Forrestal1992,Newhall2003,Ambroso2005,Katsuragi2007, Goldman2008,Nelson2008,Goldman2010,Clark2012,Clark2013,Clark2014}, and intruder impacts are common in many natural and man-made processes. In addition, since grains near the free surface are uncompressed, force transmission in this region involves inherently nonlinear effects that are of interest in their own right~\cite{Nesterenko1994,Nesterenko2001, Daraio2005,Schreck2011,Owens2011,Vitelli2012,Wildernberg2013}. That is, particularly during the important initial stages of impact, force propagation is always nonlinear. In this Letter, we describe experimental studies of the speed and spatial structure of nonlinear force propagation caused by an impact of a massive intruder into vertically oriented 2D granular beds of frictional photoelastic disks~\cite{Clark2012,Clark2014}. We specifically focus on understanding the nature of this nonlinear force propagation over a wide range of impact velocities.

For moderate-speed impacts into frictional granular materials, transmitted forces are not homogeneous. Rather, they are carried dynamically on a relatively small, quasi-linear subset of grains~\cite{Clark2012}, resembling the force chains that are common in quasi-static systems. Therefore, understanding the nature and role of the force networks is crucial to understanding force propagation. Additionally, force propagation depends on the inter-particle force law, $f(\delta)$, where $\delta$ is the particle compression. Typically $f \propto \delta^{\alpha}$, with $\alpha > 1$. The complex spatial structure of the strong force network and the nonlinear inter-particle force law mean that a macroscopic linear wave description is invalid for impacts onto free granular surfaces, or any other situation where dynamic stresses are large compared to the original confining pressure~\cite{Makse1999,Jia1999,Owens2011,Bassett2012,Schreck2011,Vitelli2012,Wildernberg2013}. For instance, the stiffness at a contact, $-df/d\delta \propto \delta^{\alpha -1}$~\cite{vanHecke2010,Siber2013,Coste1998}, approaches zero in the limit when the system is uncompressed, and thus the linear sound speed vanishes. Forces then propagate with a speed, $v_f$, which is very sensitive to $\alpha$ and the strength of the propagating forces~\cite{Nesterenko1994,Wildernberg2013,Nesterenko2001,Daraio2005,Vitelli2012}.

Several recent studies are particularly germane to this work: model studies by Gomez et al.~\cite{Vitelli2012} on shock propagation in systems of frictionless grains, model studies by Nesterenko~\cite{Nesterenko1994,Nesterenko2001} on solitons that propagate along 1D granular chains, and experiments by van den Wildenberg et al.~\cite{Wildernberg2013}. In the nonlinear limit, all of these show a force propagation speed, $v_f$, which scales as a power law in the maximum grain velocity (or the characteristic driving velocity), $v_0$:
\begin{equation}
\frac{v_f}{v_b} \propto \left(\frac{v_0}{v_b}\right)^{(\alpha -1)/(\alpha +1)},
\label{eqn:scaling}
\end{equation}
where $v_b$ is a characteristic sound speed inside a grain. 

Experiments typically differ from models described by this equation~\cite{Nesterenko1994,Nesterenko2001,Vitelli2012}, in that they involve dissipative grain interactions and force chains which are not simply lines of particles. For instance, the derivation of Eq.~\eqref{eqn:scaling} in~\cite{Vitelli2012} assumes conservative interactions, equipartition of energy between kinetic and potential, and relatively homogeneous fronts. In contrast, our particles have non-conservative interactions (non-vanishing friction and restitution coefficients less than 1), and forces propagate along complex, inhomogeneous force chains~\cite{Clark2012}, implying very rough fronts. Hence, we raise and address three questions: (1) How do friction and dissipation affect force transmission, and to what extent do predictions from frictionless models with only normal forces apply? (2) What are the key aspects that control transmission along the force networks? (3) As the relative impact speed grows, what structural changes occur in the force networks, and how do such changes affect force transmission? We note time-of-flight experiments in a 3D system in~\cite{Wildernberg2013} included friction and dissipation, and these results are consistent with Eq.~\eqref{eqn:scaling}. However, they are limited to $v_0\ll v_f$ and do not consider the spatial structure of the forces.

Our experimental apparatus~\cite{Clark2012,Clark2013,Clark2014} consists of two Plexiglas sheets (0.91~m $\times$ 1.22~m $\times$ 1.25~cm) separated by a thin gap (3.3~mm) filled with photoelastic disks (3~mm thick). Intruders are machined from bronze sheet (bulk density of 8.91~g/cm$^3$ and thickness of 0.23~cm) into disks of diameters $D$ of 6.35~cm, 12.7~cm, and 20.32~cm. We drop these intruders from a height $H \leq 2.2$~m, through a shaft connected to the top of the apparatus, yielding $v_0\simeq (2gH)^{1/2}\leq 6.6$~m/s. A Photron FASTCAM SA5 records results at frame rates of 10, 25, and 40~kHz for the soft, intermediate, and hard particles, respectively. To measure $v_0$, we track the intruder and take a numerical derivative as in~\cite{Clark2012,Clark2013,Clark2014}. The intensity of the photoelastic images yields the spatial structure of forces and allows us to measure $v_f$ (with an uncertainty of $\sim \pm5\%$).

The impact speed, $v_0$, is limited, so we use a novel approach where we vary $v_0/v_b$, and hence $v_f/v_b$, by using particles made from three photoelastic materials, each with a different stiffness (but otherwise similar). Softer grains transmit forces more slowly (i.e., smaller $v_b$), allowing us to access $v_0$ approaching typical $v_f$. Two sets of particles are made from polyurethane sheet from Precision Urethane with hardness ratings of Shore 60A (softest) and Shore 80A (intermediate), cut into disks of 6 mm and 9 mm diameter. A third set of particles is cut from the stiffest material, PSM-1, manufactured by Vishay Precision Group, which is cut into disks with diameters of 4.3~mm and 6~mm. In separate experiments, we compress individual particles between two plates (similar to particle compression in force chains) and find that a single scaling relation captures the behavior of all types of particles; see Supplemental Material (SM) for details~\cite{SuppMat}:
\begin{equation}
f = E^* w d \left(\frac{\delta}{d}\right)^\alpha,
\label{eqn:force-law}
\end{equation}
where $f$ is the compression force, $w$ is the particle thickness, $d$ is the particle diameter, $\delta$ is the displacement, and $\alpha \approx 1.4$ for all particles. The effective Young's modulus, $E^*$, is set by properties of the bulk material, including effects from the Poisson ratio and possibly other geometrical considerations~\cite{Vitelli2012}. We measure $E^* \approx 3$~MPa for Shore 60A, $E^* \approx 23$~MPa for Shore 80A, and $E^* \approx 360$~MPa for PSM-1. Additionally, when we uniaxially compress small collections of roughly 100 particles by an amount $\Delta$, we observe force chains, and we find a collective force response $F \sim \Delta^{\alpha}$, with $\alpha\approx 1.4$, for moderate $\Delta$ (see SM~\cite{SuppMat}). However, for large $\Delta$ with the softest particles, the force network is no longer chain-like, and $\alpha$ increases to $\alpha\approx 2.2$ at the maximum $F$ that we measure. This point is crucial in our analysis below.

To combine and generalize the results from all impacts, we use a simple framework centered on the collision time between grains, which contains the details of the inter-particle force law. Specifically, we show that the nature of force propagation depends on a dimensionless parameter, $M'=t_c v_0/d$, where $d$ is the particle diameter and $t_c$ is the collision time for a pair of grains impacting at relative speed $v_0$. We note that $M'$ can be interpreted as the ratio of $t_c$ to the time for the intruder to move one grain diameter, $d/v_0$ (this is similar to a dimensionless number used by Campbell~\cite{Campbell2002} in simulations of shear flows). Equation~\eqref{eqn:force-law} can be used to calculate $t_c$, and thus $M'$, as shown in SM~\cite{SuppMat}:
\begin{equation}
t_c =d v_0^{(1-\alpha)/(1+\alpha)} v_b ^{-2/(\alpha +1)} C(\alpha),
\label{eqn:tc}
\end{equation}
where $C(\alpha)$ depends on $\alpha$ but otherwise is a constant, $v_b=(E^*/ \rho)^{1/2}$, and $\rho$ is the bulk mass density of the particles. We note a numerical value of $C(1.4)\approx 4.35$. As expected for a nonlinear force law, $t_c$ depends on the relative impact speed between grains, $v_0$. Note that Eq.~\eqref{eqn:tc} implies that $M' = C(\alpha) (v_0/v_b)^{2/(\alpha+1)}$.

\begin{figure}
\raggedright (a)\\ \includegraphics[trim=20mm 0mm 20mm 0mm, clip, width=\columnwidth]{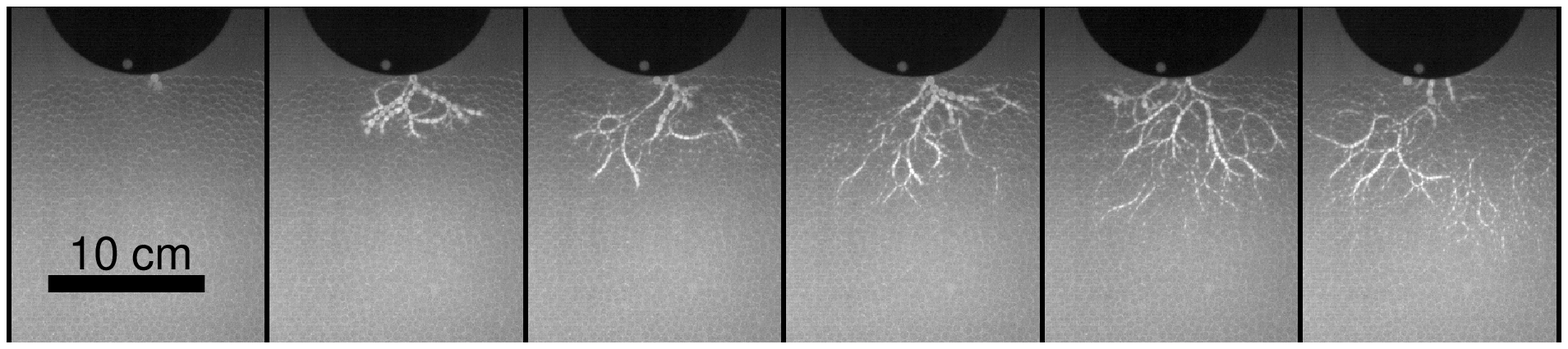}
\raggedright (b)\\ \includegraphics[trim=20mm 0mm 20mm 0mm, clip, width=\columnwidth]{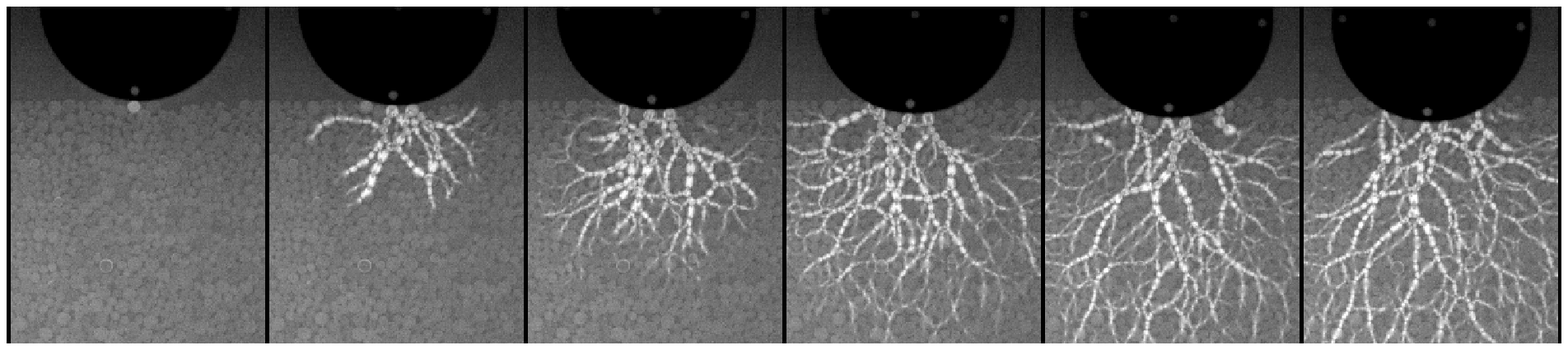}
\raggedright (c)\\ \includegraphics[trim=20mm 0mm 20mm 0mm, clip, width=\columnwidth]{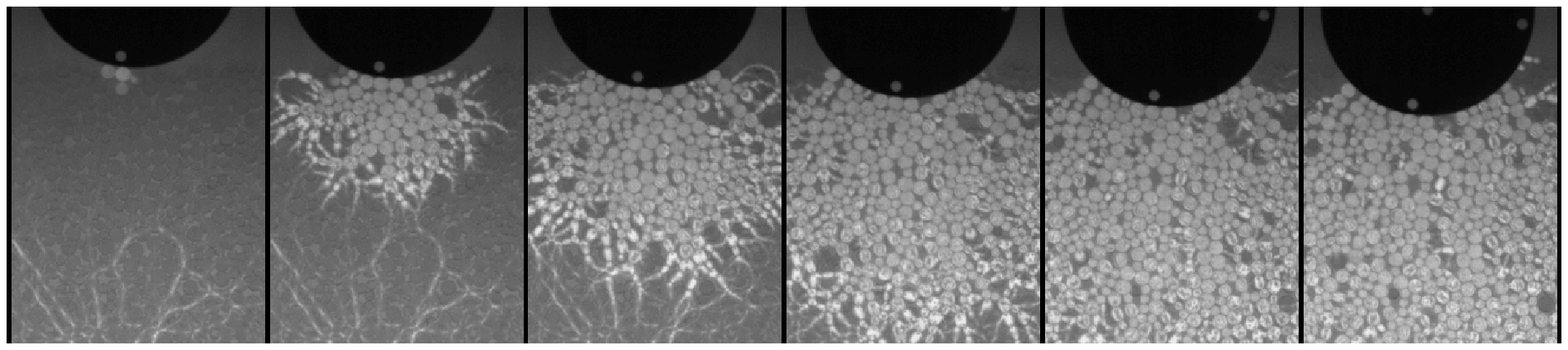}
\caption{Force propagation after impacts with $v_0\approx 5$~m/s. (a) Hard particles ($M'\approx 0.1$) show sparse, chain-like force propagation. (b) Forces for intermediate particles ($M'\approx 0.3$) are more dense spatially, but still relatively chain-like. (c) Soft particles ($M'\approx 0.6$) show a dense force structure which propagates with a well defined front.}
\label{fig:frames}
\end{figure}

At impact, the intruder makes strong contact with a subset of grains around its perimeter, and forces propagate into the material from these contacts, as in Fig.~\ref{fig:frames} and Supplemental Videos 1-3~\cite{SuppMat}. Figure~\ref{fig:frames} shows that the fraction of strong contacts and the spatial structure of the transmitted forces vary with $M'$. For small $M'$, as in~\cite{Clark2012,Clark2014}, forces are transmitted into the material by compression pulses which travel along a spatially sparse, chain-like force network. The result is a very rough front. As $M'$ increases, $M' \rightarrow 1$, the force network becomes nearly space-filling, and the front becomes smoother.

\begin{figure}
\centering
\raggedright (a) \\ \includegraphics[trim = 0mm 0mm 0mm 2.5mm, clip, width=\columnwidth]{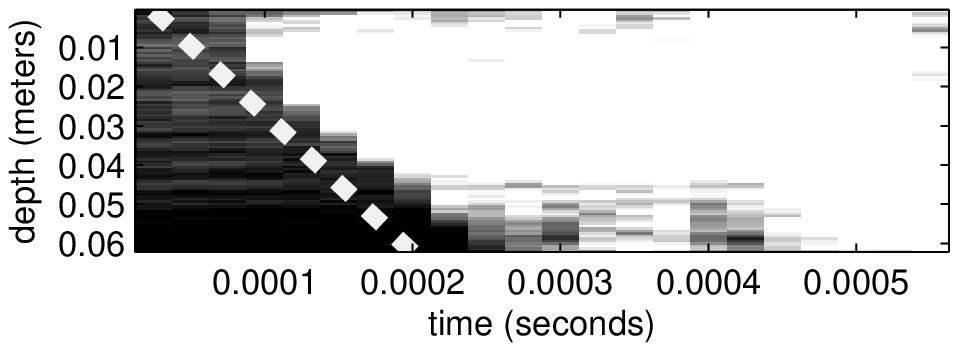}
\raggedright (b) \\ \includegraphics[trim = 0mm 0mm 0mm 2.5mm, clip, width=\columnwidth]{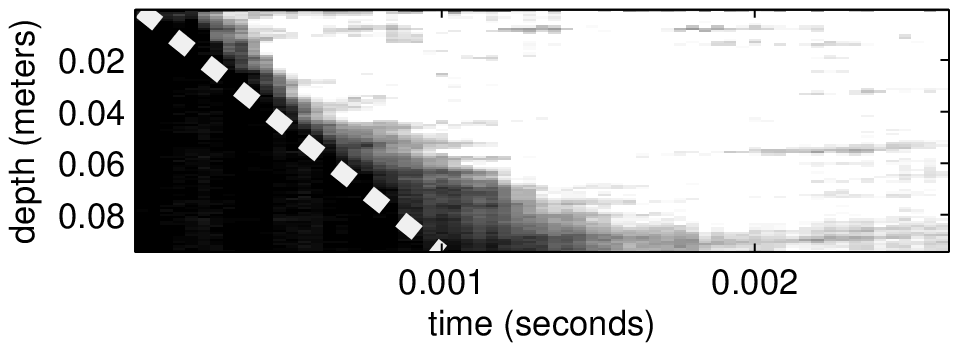}
\raggedright (c) \\ \includegraphics[trim = 0mm 0mm 0mm 2.5mm, clip, width=\columnwidth]{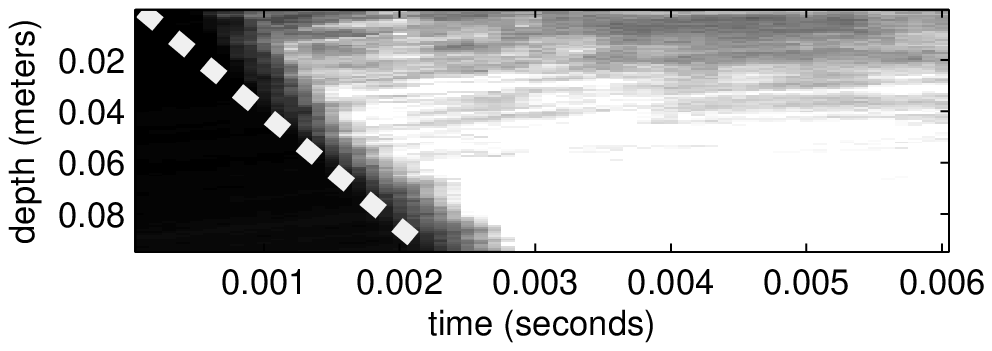}
\raggedright (d) \hspace*{33mm} (e) \\ 
\includegraphics[trim = 0mm 0mm 3mm 2.5mm, clip, width=0.45\columnwidth]{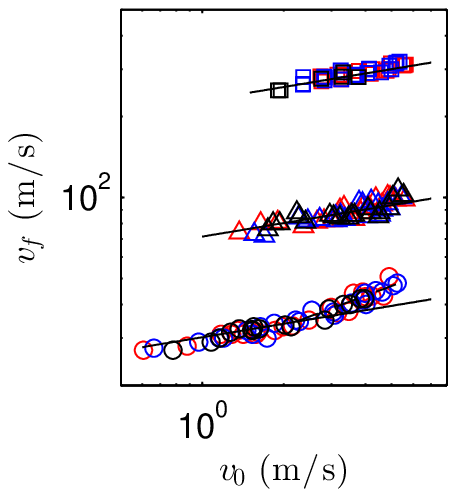}
\includegraphics[trim = 0mm 0mm 4mm 2.5mm, clip,
  width=0.45\columnwidth]{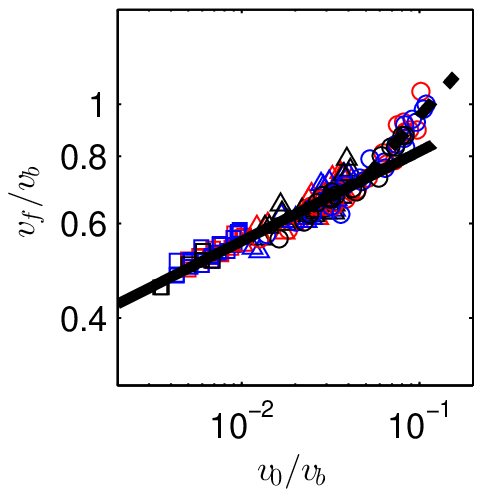}
\caption{(a)-(c) Space-time plots of propagating forces shown in Fig.~\ref{fig:frames} (see text for details). Dashed white lines indicate $v_f$. (d) These are plotted versus $v_0$. Symbol shape denotes particle stiffness (squares/triangles/circles are hard/medium/soft, respectively), and color denotes intruder diameter (red/blue/black for 6.35/12.7/20.32~cm, respectively). (e) Plot of data in (d), but with $v_f$ and $v_0$ normalized by $v_b$. The fit lines correspond to Eq.~\eqref{eqn:scaling} with $\alpha\approx 1.4$ (solid line) and $\alpha\approx 2.2$ (dashed line); see text for discussion.}
\label{fig:spacetime-plots}
\end{figure}

We determine $v_f$ with space-time plots of photoelastic images, as in Fig.~\ref{fig:spacetime-plots}(a)-(c), which shows the total photoelastic intensity as a function of time and of distance from the leading edge of the intruder. We average over an angular region spanning 90 degrees beneath the intruder, and thus coarse grain the otherwise inhomogeneous force response. White (black) corresponds to a strong (weak) photoelastic response (note, the grey region at the top of Fig.~\ref{fig:spacetime-plots}(c) represents the region just beneath the intruder in Fig.~\ref{fig:frames}(c), where particles are compressed sufficiently that the photoelastic response actually decreases slightly, due to contrasting bright and dark fringes). The slope of the leading edge in each space-time plot gives $v_f$. In Fig.~\ref{fig:frames}(d), we plot $v_f$ as a function of $v_0$ for all impacts, along with fit lines $v_f= 1.2 v_b (v_0/v_b)^{(\alpha-1)/(\alpha+1)}$, according to Eq.~\eqref{eqn:scaling}, with $\alpha=1.4$. This expression, with prefactor of $1.2$, fits all the data satisfactorily, except for the softest particles at higher velocity impacts ($v_0>3$~m/s), where a modified relation is needed. Figure~\ref{fig:spacetime-plots}(e) shows this data when all velocities are scaled by $v_b$, as in Eq.~\eqref{eqn:scaling}. The fit lines shown have exponent 1/6 (solid line) and 3/8 (dashed line), following from Eq.~\eqref{eqn:scaling}, with exponents $\alpha=1.4$ and $\alpha= 2.2$, respectively. The latter corresponds to soft particles under significant compression, as discussed previously. In SM~\cite{SuppMat}, we show data which demonstrates that the pressure during high velocity impacts into soft particles corresponds to the pressure where the exponent transitions from $\alpha\approx 1.4$ to $\alpha\approx 2.2$ in static compression tests. Thus, Eq.~\eqref{eqn:scaling} is still satisfied in this regime, provided $\alpha$ in Eq.~\eqref{eqn:tc} is replaced with $\alpha \approx 2.2$.

Figures~\ref{fig:frames} and~\ref{fig:spacetime-plots} reveal important similarities and differences with previous theoretical work. As in the simulations from~\cite{Vitelli2012}, there is shock-like behavior in our system of frictional particles, satisfying Eq.~\eqref{eqn:scaling} for impacts at low $M'$. But unlike~\cite{Vitelli2012}, these shocks are carried by a subset of particles, so that the effective front is not at all smooth. This network of force chains is roughly analogous to the 1D chains in~\cite{Nesterenko1994,Nesterenko2001}, but the network is clearly more complex than a single linear chain. For larger $M'$, new force networks form before old ones have disappeared, leading to denser networks as the intruder advances.  As $M' \rightarrow 1$ (high speed impacts into soft particles), all grains near the intruder are compressed, with a clear boundary between compressed and uncompressed grains (i.e., similar to~\cite{Vitelli2012}), but this is the regime where the scaling we observe departs from Eq.~\eqref{eqn:scaling}.

We propose that the key feature for propagation at low $M'$ is the fact that forces are carried from grain to grain along force chains in a time of order $t_c$, which is the time to load a nonlinear contact. Along a line of grains in the quasi-linear network, the speed for force transmission is simply $v_f \propto d/t_c$ (somewhat similar to an argument presented Newton's cradle experiments~\cite{Lovett1988}), which reproduces Eq.~\eqref{eqn:scaling}. Using the form from Eq.~\eqref{eqn:tc}, we obtain
\begin{equation}
v_f \propto \frac{d}{t_c} = v_0^{(\alpha -1)/(\alpha +1)} v_b ^{2/(\alpha +1)} \left[C(\alpha)\right]^{-1},
\label{eqn:v-scaling}
\end{equation}
which in turn yields Eq.~\eqref{eqn:scaling}. We emphasize that this scenario does not require energy conservation, and it does not require simple linear chains. For instance, even with restitutional losses during inter-particle interactions, $t_c$ will still have a similar form to Eq.~\eqref{eqn:tc} if the contact force is initially zero or very small. This expression can be applied in related systems, given a force law, even for spatially complex, nonconservative forces.

\begin{figure}
\raggedright (a) \\
\centering \includegraphics[trim = 2mm 0mm 0mm 2.5mm, clip, width=0.9\columnwidth]{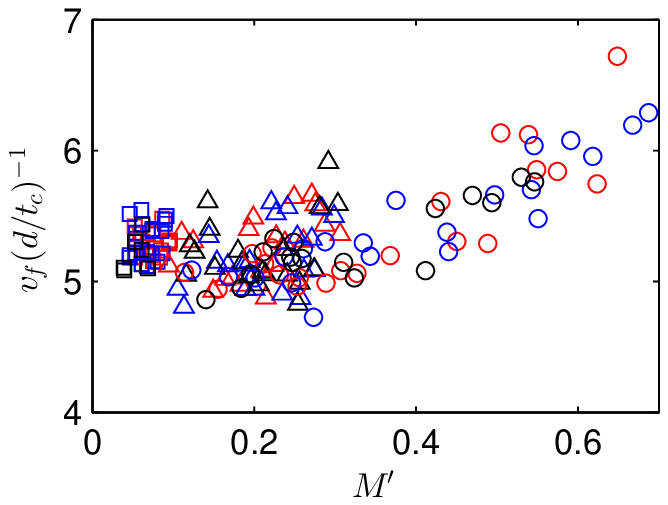}
\\ \raggedright (b) \\ 
\centering \includegraphics[trim = 0mm 0mm 0mm 2.5mm, clip, width=0.9\columnwidth]{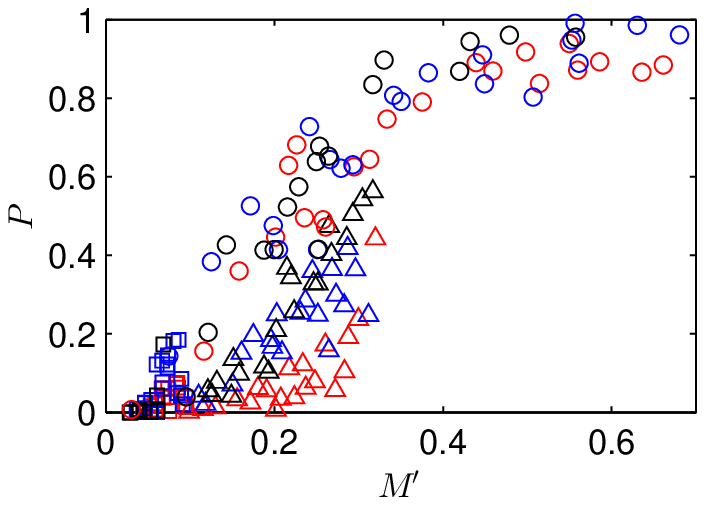}
\caption{(a) Plot of ratio of measured $v_f$ to $d/t_c$ versus $M'$. We find $v_f\approx 5.3(d/t_c)$ to hold for small $M'$, but for large $M'$, $v_f$ is faster than this relation. (b) Participation ratio, $P$, versus $M'$, which approximates the spatial density of propagating forces (see text for details).}
\label{fig:M-prime-scaling}
\end{figure}

Figure~\ref{fig:M-prime-scaling} shows a summary of our data in the context of the collision time framework. Figure~\ref{fig:M-prime-scaling}(a) shows the data from Fig.~\ref{fig:spacetime-plots}, plotted as a function of $M'$. For $M' \stackrel{<}{~} 0.4$, we find $v_f\approx 5.3 (d/t_c)$, whereas at larger values of $M'$, $v_f$ increases faster than predicted by this scaling relation. The spatial density of the forces also changes with increasing $M'$, as seen in Fig.~\ref{fig:frames}, and the front becomes much smoother. To quantify this and compare it to the departure from $v_f \propto d/t_c$, we define a participation ratio $P$ as the fraction of grains beneath the intruder exhibiting a strong force. To estimate $P$, we examine all images between $4.5 t_c$ and $5.5 t_c$. We threshold each image into bright and dark pixels, such that bright pixels are a fixed percentage greater than the background intensity: 25\% for hard particles, 30\% for intermediate stiffness particles, and 45\% for the soft particles. These values are chosen such that the resulting black-and-white image appears most similar to the original photoelastic image (they are slightly different for each material, due to differences in photoelastic response, camera settings, and lighting conditions). We calculate the average density of bright pixels in a quarter annulus extending downward from the intruder by $25d\approx 15$~cm, corresponding to the distance the front would travel in $5t_c$. This value is averaged from $4.5 t_c$ to $5.5 t_c$ and divided by the approximate packing density of 0.8. Figure~\ref{fig:M-prime-scaling}(b) shows the resulting $P$ plotted versus $M'$. Despite some expected scatter due to different particle types and fluctuations between experiments, we observe that between $M'=0.2$ and $M'=0.4$, $P$ rises from near 0 to almost 1. Thus, the collective stiffening ($\alpha \approx 2.2$) observed with increasing $M'$, which causes the departure from Eq.~\eqref{eqn:scaling}, is also correlated with significant homogenization of the strong force network.

We note two possible grain-scale sources for these effects. First, spatially dense force networks tend to have more force contacts per particle. In this case, a pair-wise Hertz-like force law may fail, since deformation at one contact can affect the response at other contacts. Second, lateral expansion of grains forms new contacts, leading to a collective strengthening of the system. These effects have been demonstrated recently in a 3D system of soft particles~\cite{Brodu2015}. In SM~\cite{SuppMat}, we show that this occurs for our softest particles. Photoelastic images in SM~\cite{SuppMat} from compression tests show only chain-like forces for hard and medium particles and spatially dense forces in the softest particles at large compression. These images demonstrate particles deforming and forming additional contacts in the lateral direction. Other studies have observed this effect when physical grains with a nonzero positive Poisson ratio are compressed sufficiently~\cite{Siber2013,Owens2011}. When compression forces propagate along relatively sparse force networks, the compressed particles can easily expand laterally, with no resistance from neighboring non-force-chain particles. However, larger stress (or higher $M'$) leads to a denser strong force network and a significant number of new contacts in the lateral direction, which stiffen the material. This situation may occur in many common non-brittle materials, including some soils, food grains, plastics, rubber, etc., although brittle grains may break for comparable strains. 

We conclude by returning to the questions raised in the introduction. We find Eq.~\eqref{eqn:scaling} to hold for impacts into frictional particles over a substantial range of impact speeds. The spatial structure of the networks along which forces are carried appears to differ significantly from what was reported for frictionless simulations in~\cite{Vitelli2012}. For low $M'$, we observe strong force networks that are spatially sparse and inhomogeneous. The networks become more spatially dense and homogeneous (i.e., $P$ grows) and the data for scaled $v_f$ depart from Eq.~\eqref{eqn:scaling} as $M'$ increases. This departure is associated with a collective stiffening also observed in static compression tests, which can quantitatively account for departure from Eq.~\eqref{eqn:scaling}. These effects arise from the lateral expansion of physical grains, an effect which is necessarily absent in DEM/MD simulations for which there is no interdependence of multiple contact forces acting on a given grain.

We thank J. Dijksman for helpful discussions, and we acknowledge funding from DTRA, under Grant No. HDTRA1-10-0021.

\bibliography{References}

\end{document}